\documentclass[showpacs,prc]{revtex4}

\usepackage{graphicx}
\usepackage{dcolumn}
\usepackage{bm}

\begin{document}
\title{Theories of Low Energy Nuclear Transmutations}
\author{Y. N. Srivastava}
\affiliation{Department of Physics \& INFN, University of Perugia, Perugia, IT}
\author{A. Widom and J. Swain}
\affiliation{Physics Department, Northeastern University, Boston MA, USA}

\begin{abstract}
Employing concrete examples from nuclear physics it is shown 
that low energy nuclear reactions can and have been induced 
by all of the four fundamental interactions (i) (stellar) gravitational, (ii) strong, 
(iii) electromagnetic and (iv) weak.  Differences are highlighted through 
the great diversity in the rates and similarity through the nature of the nuclear 
reactions initiated by each.  
\end{abstract}


\maketitle

\section{Introduction \label{intro}}

We show below through physical examples that low energy nuclear reactions have been 
induced by {\em all} of the four fundamental interactions: gravitational, strong, 
electromagnetic and weak. 
\medskip
\par \noindent
{\bf Gravity}:  Gravitational interactions are well known to cause nuclear reactions 
and fusion in a star. Were it not for nuclear activity, a star would be dynamically 
unstable and undergo gravitational collapse\cite{Bethe, BetheNobel}. In fact, standard 
theory predicts the collapse of a star when the nuclear fuel is exhausted as the star 
can no longer counteract the inward compression due to gravitation. 
\medskip
\par \noindent
{\bf Nuclear}: A laboratory example of low energy strong interaction fusion is provided 
by a fast discharge in fine deuterated polymer fibers. In such fibers, deuterons are 
accelerated to speeds  high enough to overcome the barrier due to mutual 
Coulomb repulsion, giving rise to the production of $2.5\ MeV$ neutrons through low 
energy reactions such as 
\begin{equation}
\label{W1}
d + d \rightarrow  n +\ ^3{\rm He}.
\end{equation}   
In the same set of experiments\cite{Stephanakis}, also {\it non deuterated} fibers 
exhibit an ``anomalous'' production of neutrons, at a rate over 6 orders of magnitude 
larger than that expected through natural contamination of deuterons in a normal material. 
Such experimental evidence strongly suggests an explanation in terms of weak interaction 
induced nuclear reactions\cite{Pramana}. This will be discussed in a later 
Sec.{\ref{Nondeuterated}}.
\medskip
\par \noindent
{\bf Electromagnetic}: Purely electromagnetically induced fusion has been observed through the 
Coulomb explosion of large deuterium clusters -stripped of their electrons by lasers- and 
their subsequent blow up resulting in nuclear fusion\cite{Zweiback}. In more recent experiments, 
other charged nuclei have been exploded and the ensuing nuclear reactions have been observed. 
\medskip
\par \noindent
{\bf Weak}: In the corona of the Sun, magnetic fields emerge from one sunspot to dive into 
another. The coronal magnetic field accelerates the electrons and protons enough to cause 
production of neutrons through weak interactions\cite{WSL}. These neutrons then cause further 
nuclear transmutations of the available material accounting for the presence of 
anomalously formed nuclei on the surface of the Sun and in the Solar 
corona\cite{Wurz, Cowley, Goriely, Lodders}. 

Once in a while, magnetic flux tubes explode leading to spectacular solar flares releasing 
extremely high energy particles some of which reaching the earth. Often, the resultant 
electromagnetic fields are large enough to disturb terrestrial satellite communications and 
secondary muons produced in the upper atmosphere in sufficient numbers to have been detected 
in the underground CERN L3+C detector\cite{Pramana,L3}.

In the following work, we shall analyze in some detail, various mechanisms and phenomena found both 
in Nature and in the laboratory related to low energy nuclear reactions. We shall  make a 
general classification of these diverse phenomena in order to delineate common themes (such 
as collectivity, thermal or non-thermal nature of their induction) as well as profound 
differences (such as the time scale for induction by a particular interaction) between nuclear 
reactions caused by different interactions under diverse circumstances. We shall also illustrate 
the interplay between various interactions necessary for the genesis of a given physical process. 
For the energy production in a star, all four interactions are essential.  

\section{Gravitational LENT \label{GLENT}}

\subsection{From Helmholtz  to Bethe \label{HB}}
In the standard model of star formation, collectivity plays a central role. It is asserted 
that undamped density fluctuations cause an astronomically large number of interstellar gas 
(essentially hydrogen) to clump together, whereupon gravity begins to compress them further 
thereby increasing the protostar's temperature. 

To appreciate the subtle interplay existing between different interactions in causing nuclear 
reactions, the leitmotif of our paper, it is useful to recall the paradigm shifts in the view 
that earlier physicists had to make regarding the main source of a star's energy production and 
stability, as recounted by Hans Bethe\cite{BetheNobel}, a leading architect of gravitational low 
energy nuclear theory (Gravitational LENT).

To estimate the stability of a star, say our Sun, Helmholtz -around the 1850's- was the first 
to use the only tool available to him: Newtonian gravity. If one gram of matter falls on the 
Sun's surface, it acquires a potential energy $(\Delta E)_{pot}$ given by  
\begin{equation}
\label{G1}
\frac{\Delta E_{pot}}{\rm gm} 
= - \frac{GM}{R} = - 1.91 \times  10^{15}\ \frac{\rm erg}{\rm gm}\ .
\end{equation}
A similar energy must have been set free when the Sun was assembled. Through the virial theorem, 
he argued, one half of it must have been set free as kinetic energy [which allows us to estimate 
the temperature] and the other half must have been radiated away. At present, the outward flux 
of radiation from our Sun is 
\begin{equation}
\label{G2}
{\cal F}_{radiation} = 1.96\ \frac{\rm erg}{\rm gm\ sec}\ .
\end{equation}
Hence, it was estimated that if gravitation supplies the energy, then in 
\begin{math} 10^{15}\ {\rm sec}\approx 30{\rm \ million\ years} \end{math} the Sun would have 
radiated away all its energy and would have undergone a gravitational collapse. But Darwin, other 
biologists and geologists argued with Helmholtz (and Kelvin\cite{Kelvin}, who concurred with Helmholtz) that 
they  needed a much longer life time for the Sun and that his gravitational mechanism for the 
source of energy production in the Sun must be in error. Of course, they were right. In fact, 
as we will see, gravitation supplies the necessary force to hold and compress the particles 
collectively and raise their temperature for positively charged nuclei to overcome the Coulomb 
barrier but it is the weak force which ignites the nuclear fire by providing neutrons without 
which no deuterons or heavier nuclei could be formed.   

In 1895, radioactivity in the weak sector of the standard model of 
fundamental interactions would be discovered by Henri Becquerel, Pierre Curie and Marie Curie 
and through it the age of the Earth  and later that of meteorites and other geological objects 
would be determined. One deduces the age of the Sun from that of the meteorites to be 4.5 billion 
years in close agreement with the age 4.56 billions deduced from Solar oscillations\cite{AgeSun}. 
Clearly, we need some interaction other than gravitation to fuel the Sun.   

Since the Sun has existed for at least 4.5 billion years, we may deduce that the gain in energy 
per gm to be provided by whatever agency must be at least  
\begin{equation}
\frac{\Delta E_{gain}}{\rm gm} \gtrsim\ 3 \times 10^{17}\ \frac{\rm erg}{\rm gm}\ .
\label{G3}
\end{equation}
If not gravitation, then which of the other three EM, weak or strong can be the source? If for simplicity 
we assume that initially we only had electrons and protons and we want to fuse the protons into 
heavier nuclei, then we need the intervention of neutrons. Electromagnetism can not do it by itself because 
it can not change the charge. Also, the direct strong interaction between two protons to change their 
charges is well nigh negligible since  the temperature of the core is estimated to be about 
\begin{math} T_{core}\approx 17\ {\rm million\  K} \end{math}. This corresponds to a mean thermal energy 
for the electrons and protons to be a mere \begin{math} 1.46\ {\rm KeV}\end{math}. Hence, the only 
alternative left is to invoke the weak interaction. The Fermi theory does provide 
a possibility of neutron production through the inverse beta decay   
\begin{equation}
\label{G4}
e^- + p^+ \rightarrow n + \nu_e
\end{equation}
This reaction was first proposed by Wick and later independently by Yukawa and Sakata for 
the observed phenomena of electron capture from the K and L shells from a large proton rich nucleus. 
But notice that in an electron capture, the electron is bound near the nucleus, the proton is bound within the 
nucleus and the subsequent neutron remains bound to the nucleus. All that escapes is the unobserved 
neutrino. Of course, this is not quite exact. As the closest to nucleus electron disappears, a higher 
shell electron jumps down to replace it, a third to displace the second, etc., thus leading to a 
cascade of emitted photons. It is precisely through the emitted photons, chiefly in the X-ray region, 
that the process can be directly registered at all.  Alvarez\cite{Alvarez} provided the first clinching experimental 
evidence that the \begin{math} ^{67}{\rm Ga} \end{math} nucleus captures one of its 
orbital electrons and converts to the stable nucleus \begin{math} ^{67}{\rm Zn} \end{math}.

The life time of K capture and hence the transformation of one atom to a different atom can be of the 
order of seconds to months depending upon the mass difference between the parent and the daughter atom. 
It is well to remember that even in the vacuum, prior to the intervention of temperature, several orders 
of magnitude difference in the rates  can occur due to differences in phase space.

Instead, for the reaction in Eq.(\ref{G4}), both the electron and the proton have negligible kinetic 
energy on the nuclear \begin{math} {\rm MeV} \end{math} scale. Hence, the mean \begin{math} Q \end{math} of the 
reaction is negative: \begin{math} Q \approx - 0.78\ {\rm MeV} \end{math}. One may consider the thermally 
induced process
\begin{equation}
\label{G5}
W_{\rm thermal} + e^- + p^+ \rightarrow\ n + \nu_e,
\end{equation}
where the thermal radiation \begin{math} W_{\rm thermal} \end{math}
supplies kinetic energy to the initial system. But the available thermal  energy even near the core 
of the Sun  \begin{math} \sim KeV \end{math} is miniscule in comparison with the threshold energy 
\begin{math} \sim 0.78 {\rm MeV} \end{math} required for the reaction to proceed. Hence, the process 
can occur through thermal fluctuations albeit with an exceedingly small probability. 

The dramatic difference between the time scales involved between this case, Bethe's estimate for the 
core of the Sun is \begin{math} 10^{30}{\rm \ years} \end{math}, and the observed K captures, 
typically a few days, is due to the large exponential threshold barrier in the first case. This will  
be of crucial interest for us in what follows, where the reaction in Eq.(\ref{G4}) would be invoked 
for specially designed condensed matter systems in which the electron kinetic energy can be boosted 
enough or in a relativistically covariant language, the electron mass can get renormalized by external 
electric fields radiated from currents, 
\begin{equation}
\label{G6}
W_{\rm EM} + (N+1) e^- + p^+ \rightarrow n + N e^- + \nu_e ,
\end{equation}
so as to overcome the threshold barrier \begin{math} Q \end{math} value turning positive, as in the 
electron capture case, with substantial rates for neutron production. In Eq.(\ref{G6}), anticipating 
effects due to electron collectivity, we have generalized the reaction in Eq.(\ref{G4}) to the case of 
\begin{math} N + 1 \end{math} initial electrons which end up with \begin{math} N \end{math} final 
electrons after the weak destruction of a single electron. As we shall see in Sec.(\ref{Betatron}), the 
presence of a large number of electrons in dense matter systems can, through the Darwin interaction, 
change the energy distribution of a single electron. In such cases, the rates for the reaction 
in Eq.(\ref{G6}) can be quite different from that of Eq.(\ref{G5}) or for electron capture.

\subsection{Two Proton Channel Rates \label{2P}}

\begin{figure}
\scalebox {0.6}{\includegraphics{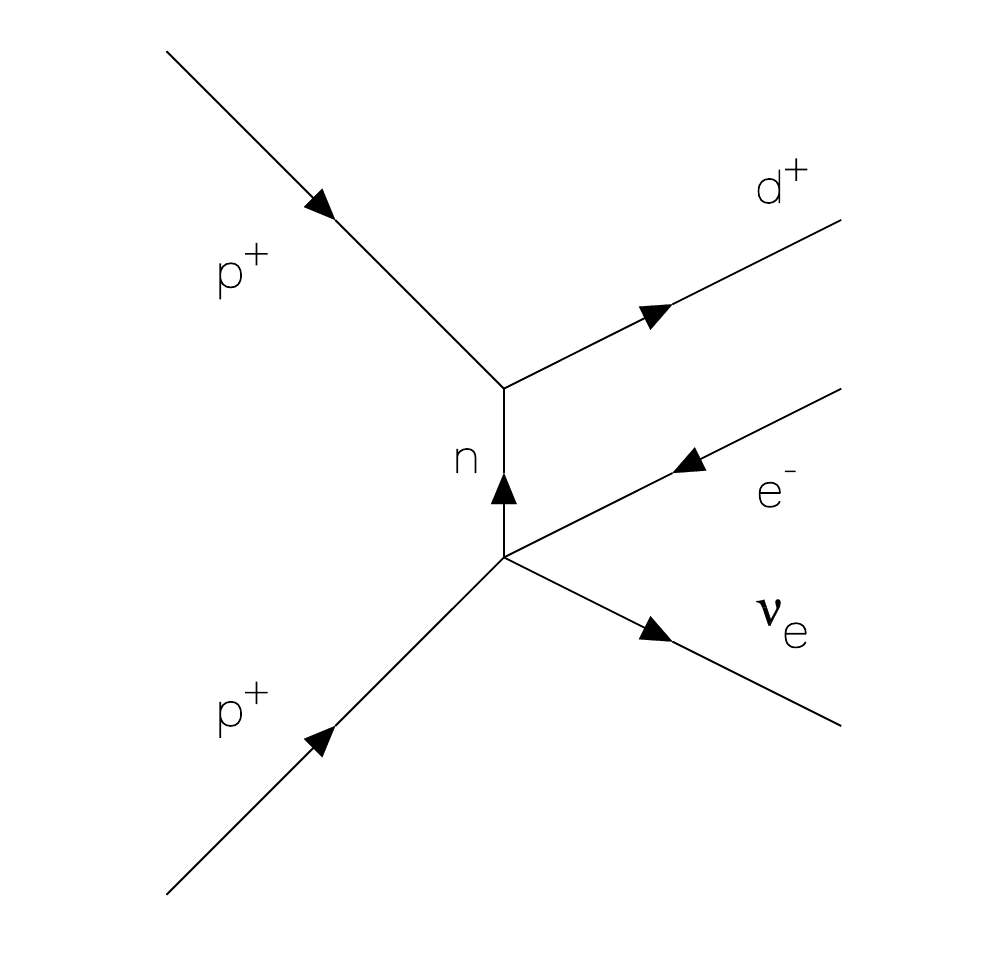}}
\caption{Shown is the Feynman diagram for the reaction 
$p^+ +p^+ \to d^+ +e^+ +\nu_e $ as in Eq.(\ref{K}).} 
\label{fig1}
\end{figure}

Given the depressingly low rate due to the threshold barrier for the direct production of neutrons 
through Eq.(\ref{G5}), von Weizs\"acker \cite{Weizsacker} suggested the exothermic reaction
\begin{equation}
\label{K}
p^+ + p^+ \rightarrow  d^+ + e^+  + \nu_e,
\end{equation}
with a gain of energy $\sim  0.4\ MeV$ from a transmutation of two protons. For this reaction, 
we can estimate the energy gain/gm to be 
\begin{equation}
\label{K1}
\frac{\Delta E_{gain}}{gm} \approx\   1.2 \times\ 10^{17} \left(\frac {erg}{gm}\right),
\end{equation}
a value quite close to that needed in Eq.(\ref{G3}) in order to keep the Sun warm. 

The completion of the mechanism described through the above processes requires the reaction 
rates for this ``$p^+p^+$'' channel of energy production. It was first computed by Bethe and 
Critchfield\cite{Bethe} and it involves all three, EM, weak and strong interactions of the 
standard model of particle physics. 
At low relative kinetic energy between the two protons, we have a strong Coulomb repulsion inhibiting
the entire process Eq.(\ref{K}). This is a two step process: first we have  the essentially ``zero'' 
range weak Fermi process $p^+_1\to \tilde{n}+e^++\nu_e$ followed by the very short range 
strong interaction process $p^+_2 + \tilde{n}\to d$, visualized in the Feynman diagram in 
FIG. \ref{fig1} .
\begin{equation}
\label{K2}
p^+_1 + p^+_2 \to (\tilde{n} + e^+ + \bar{\nu}_e) + p^+_2 \to  d + e^+ + \bar{\nu}_e, 
\end{equation}
where $\tilde{n}$ is an intermediate ``virtual'' state neutron, which combines with the second proton 
to fuse into a deuteron.

With an eye towards later usage for other processes, let us decompose the computation into (i) a repulsive 
Coulomb Gamow suppression factor $B$ and (ii) a weak-strong cross section for a generic process  
\begin{equation}
\label{K3}
N_1(Z_1) + N_2(Z_2) \rightarrow\ X,
\end{equation}
where $Z_{1,2}$ refer to the charge of the nuclei $N_{1,2}$ and $X$ refers to some specified final state. 
The effective cross section reads
\begin{equation}
\label{K4}
\sigma_{effective} = B \sigma(N_1+N_2\to\ X),
\end{equation}
where
\begin{equation}
\label{K5}
B = \frac{2\pi \gamma}{e^{2\pi\gamma} - 1}
\ \ \  {\rm and} \ \ \  \gamma = \frac{Z_1Z_2\alpha}{v/c}\ .
\end{equation}
In the above equation, $v =\ (\sqrt{2E/\mu})$ denotes the relative velocity, $\mu$ is the reduced mass, 
$\alpha =e^2/\hbar c \approx\ 1/137\ $ is the fine structure constant and $E$ is the relative kinetic energy 
when the two nuclei are far apart. The cross section $\sigma(N_1+N_2\to\ X)$ is in 
principle computable from first principles of the known weak, EM  and strong interactions. 
For low velocities, $\gamma>>1$ and hence $B$ is approximately 
\begin{eqnarray}
\label{K6}
B \approx 2\pi \gamma e^{-2\pi \gamma} = \sqrt{\frac{E_g}{E}} e^{-\sqrt{E_g/E}} 
\nonumber \\ 
E_g = (2\pi Z_1Z_2\alpha)^2 \mu c^2
\end{eqnarray}
Since, the reactions Eq.(\ref{K3}) of interest are exothermic for which the cross sections 
$\sigma\propto 1/v$ for small $v$, it is convenient to write the effective cross section as 
\begin{equation}
\label{K7}
\sigma_{effective}(N_1+N_2\to X;E) = \frac{S(E)}{E} e^{-\sqrt{E_g/E}},
\end{equation}
where $S(E)$ is a characteristic of the nuclear reaction under investigation. 
Bethe-Critchfield\cite{Bethe} obtain for deuteron production through the PP chain Eq.(\ref{K}) 
\begin{equation}
\label{K8}
S(pp\to de^+\bar{\nu}_e) = 3.36 \times\ 10^{-25} MeV\  barn,
\end{equation}
which is exceedingly small in comparison to purely nuclear reactions, i.e, where $X$ does not contain 
leptons and so no weak processes are involved. In fact, the reaction Eq.(\ref{K}) has never been 
observed in any earth laboratory. By contrast, for non-leptonic initial and final states, one finds\cite{Fowler}
\begin{equation}
\label{K9}
S_{nuclear} \sim\ (1 KeV\  barn)\div\ (10 MeV\  barn). 
\end{equation}
Typically ($1\ KeV\  barn$) results when the transition is radiative, i.e. whenever there is a photon 
in the final state whereas for purely nuclear transitions we find numbers in the ($MeV\ barn$) range.

Let us pause to understand why the reaction Eq.(\ref{K8}) has a more than $10^{22}$ lower probability 
to occur than a typical strong reaction Eq.(\ref{K9}). Since the former is a weak process with a cross 
section proportional to $(G_FE)^2$, $G_F$ being the Fermi coupling constant and $E\sim MeV$, we expect 
a suppression factor of about $10^{-16}$ with respect to the strong process. The extra suppression by 
about $10^{-6}$ arises due to the reduced probability [$\sim (E/M)^2$, with $M$ the nucleon mass] for 
a virtual neutron to combine with a proton to form a deuteron. 

Of course, Eq.(\ref{K7}) is not the end of the story. It would if it were a {\em beam} experiment, that is  
if the incident energy $E$ were fixed, But there is a Maxwellian velocity distribution which has to be
 integrated upon. Hence, Bethe and later Bahcall and others inserted it and the rate of nuclear reaction 
 at a given temperature $T$ is proportional to
\begin{equation}
\label{K7-1}
\left[\frac{8}{\pi (k_B T)^3M}\right]^{1/2}  \int dE E \sigma_{eff}(E) e^{-E/(k_BT)} .
\end{equation} 
This integration changes the final temperature or equivalently the mean energy dependence 
considerably in the rate of the reaction. One further point regarding $S_{nuclear}$ needs be made. 
Further variations can and do occur if there are resonances. This is discussed below.  

For room temperature non-thermal induction of neutron production such as electron capture in a plasmon polariton
set up, there is no cross section only a capture rate. To calculate a cross section from a rate one 
requires two beams so that one may define a cross section as a rate divided by a flux. In the context 
of computing rates when the beams are in reality oscillating plasma oscillations, one is perhaps better 
off simply quoting the rates of processes. In the Fermi theory of point interaction, the electron has 
to be on top of the nucleus for the reaction to go. The method suggested\cite{WL} employs the  process 
of radiation induced electron capture by protons or deuterons to produce ultra low momentum neutrons 
and neutrinos. In the neighborhood of metallic hydride cathodes, protons or deuterons can do such 
inverse beta decay conversions through the collective plasma modes of the charged particles. Hence, 
its proper description is through Eq.(\ref{G6}) and not through a two body cross section since electrons 
and protons or deuterons  oscillate back and forth at a frequency $\Omega$ and have a root mean square 
lateral displacement 
$\sqrt{<{\bf u}^2>}$ and when a renormalized invariant mass $m^*> \gamma_o m$ with $\gamma_o> 2.5$ is 
reached for an initial electron, it can cause the Fermi inverse beta reaction. This is strikingly 
different from a beam situation. Failure to do so and analyze this reaction using two body 
cross section at a fixed velocity would lead to quite erroneous results.

\subsection{Production of $\alpha$ Particles via Four Protons \label{alpha}}

The deuterons produced from the PP reaction can in turn produce $^3{\rm He}$ and an $\alpha$: 
\begin{equation}
\label{A1}
d + d \to\  ^4{\rm He} + \gamma
\end{equation}
\begin{equation}
\label{A2}
d + d \to\ n + ^3{\rm He}
\end{equation}
\begin{equation}
\label{A3}
p + d \to\ ^3{\rm He} + \gamma
\end{equation}
As discussed in the Sec.\ref{Tokomak} barring selection rules, $S$ for reactions Eq.(\ref{A1}) 
and Eq.(\ref{A3}) should be $\sim KeV\  barn$, whereas for Eq.(\ref{A2}) it should be 
$\sim MeV\  barn$. Also, as mentioned in Eq.(\ref{W1}),  the neutron in Eq.(\ref{A2}) has an energy 
of $2.5\ MeV$ which has been measured in exploding wire experiments\cite{Stephanakis} as well as 
in the Coulomb explosion experiments\cite{Zweiback}. Schwinger\cite{Schwinger} considered the 
radiative reaction Eq.(\ref{A3}) and observed that at low energies, if one includes just 
the $S$-wave, the dipole radiation would not contribute because of the positive intrinsic parities 
of the $p$, $d$ and $^3He$ and negative parity of $\gamma$ and hence it should be highly suppressed. 
The same argument should also apply to the radiative decay Eq.(\ref{A1}). On the other hand, all the 
intrinsic parities are positive in Eq.(\ref{A2}) and hence this process can proceed via $S$-wave. 
It is thus encouraging that both in exploding wire and Coulomb explosion experiments, reaction 
Eq.(\ref{A2}) has been observed and to our knowledge reaction Eq.(\ref{A1}) not. It is suppressed 
at the level of $10^{-7}$ in the vacuum.

Proceeding on with the PP chain, it has become customary in the astrophysics literature to invoke 
the Bethe-von Weizs\"acker weak process twice and write it as a four initial proton process
\begin{equation}
\label{A4}
p + p + p + p \to\ ^4{\rm He} + e^+ + e^+ + \nu_e + \nu_e.
\end{equation}
Needless to say that there is no possibility to ever measure this reaction directly on earth.

On the other hand, for a partial verification of the PP chain in the Sun, the detection of neutrinos arising from the fast part of the 
cycle was undertaken. As discussed in the next section, nuclei between helium and carbon are unstable. One such unstable process 
which concerns bound and free electron capture from a $^7\rm{Be}$ nucleus leading to $^7\rm{Li}$ nucleus and emission of an electron 
neutrino has been widely investigated\cite{Bahcall}. The electron capture reaction
\begin{equation}
\label{A5}
e^- + ^7{\rm Be} \to\ ^7{\rm Li} + \nu_e,
\end{equation}   
half-life has been measured in the laboratory\cite{Selove}
\begin{equation}
\label{A6}
\tau[^7{\rm Be}(e^-,\nu_e)^7{\rm Li}] =\ (4.61 \pm\ 0.001) \times 10^6\ sec.,
\end{equation}   
leading to a rate $(2.169\pm\ 0.005)\times 10^{-7}\  Hz$. This implies, that after $1\  sec.$, we 
should find about $2.17\times 10^{14}\ ^7{\rm Li}$ atoms and the same number of neutrinos if we had 
an initial supply of say $10^{22}$ $^7\rm{Be}$ atoms. Historically, the inverse electron capture 
reaction
\begin{equation}
\label{A7}
^{37}{\rm Cl}(\nu_e, e^-) ^{37}{\rm Ar}
\end{equation}   
was employed for the Solar neutrino detection on earth by Davis and others.

\subsection{CNO cycle and beyond \label{CNO}}
We mention the CNO cycle employed in massive stars simply to direct attention to the role of 
resonances. Early workers \cite{Weizsacker1, Bethe1} had difficulty in bridging the gap between 
$^4He$ and $^{12}C$ in the stars since the elements in this gap are unstable. Given the great 
abundance of helium, the process of fusing 3 alpha particles was invoked
\begin{equation}
\label{C1}
^4{\rm He} +\ ^4{\rm He} +\ ^4{\rm He} \to \ ^{12}{\rm C} + \gamma.
\end{equation}
Since it requires the simultaneous presence of three particles in the initial state, the 
probability should be extremely small except that this process is blessed with a double resonance. Salpeter\cite{Salpeter} 
pointed the first resonance which is the unstable nucleus 
$^8{\rm Be}$ with approximately the same mass as two alpha particles and  Hoyle\cite{Hoyle} noted 
the second i.e.,  $^8{\rm Be}$ and $^4{\rm He}$ have the same mass as an excited state of 
$^{12}{\rm C}$. Of course, the reaction Eq.(\ref{C1}) can not be observed in the laboratory, 
but the resonances can and have been observed.

Bethe first emphasized the great importance of the above process which should be the main reaction producing stable Carbon, 
active when the temperatures reach around $1.1\times\ 10^8\ Kelvin$. Carbon is of course considered crucial for the buildup of 
elements. Around the same temperature, another capture of an alpha particle produces oxygen
\begin{equation}
\label{C2}
^{12}{\rm C} + ^4{\rm He} \to\ ^{16}{\rm O} + \gamma.
\end{equation}
We shall not follow the astrophysical consequences any further, except to remind the reader of the 
intricate interplay between all the four interactions in producing various low energy reactions 
leading to very different rates.   

\section{Strong LENT \label{LENT}}
We now turn our attention to laboratory examples where strong interaction fusion reactions, such as
\begin{equation}
\label{S1}
d + d \to n +\ ^3{\rm He},
\end{equation}
have been directly observed through clean signals of threshold  $2.5\  MeV$ neutrons. As discussed 
in the last section, the Coulomb barrier is quite large at low energies, see Eq.(\ref{K7}), and 
thus it is well known that in the vacuum this process would have a vanishingly small probability. 
In the core of the stars, when the temperature becomes  sufficiently high to overcome the barrier, the 
process is supposed to occur. In reality, the results of such experiments do not yet have practical 
importance. An alternative method  using fast electrical discharge through thin deuterated polymer 
fibers has been very successful in causing fusion in many laboratories. This method is not directly 
thermal but uses an electrical impulse to accelerate the deuterons enough to overcome the Coulomb 
barrier. We shall discuss it in the next subsection using an example of a particular experiment.  

\subsection{Deuterated wires and neutron production \label{Deuterated}}
Let us focus on the exploding wire experiment by Stephanakis {\it et al}\cite{Stephanakis}. Their experimental setup is 
summarized below:
\medskip
\par\noindent
{\rm Fibers\ used: Polyethylene\ and\ Polypropylene}
\par\noindent
{\rm Density\ of\ deuterium} $>\ 10^{19}/cm^3$
\par\noindent
{\rm High\ power\ discharge} $\geq\ 10^{12}\ Watts$
\par\noindent
{\rm Typical\ discharge\ currents} $\sim  1.2\ Mega Amperes$
\par\noindent
{\rm  Peak\ Voltage} $\sim 600\ Kilo Volts$
\par\noindent
{\rm Duration\ of\ the\ discharges} $\sim\ 50\ nanoseconds$
\par\noindent
{\rm Length\ of\ the\ fibers} $\sim  3.5\ cm$
\par\noindent
{\rm Diameter\ of\ the\ fibers} $10\ to\ 190\ microns$
\par\noindent
{\rm Mean\ kinetic\ energy\ of\ the\ electrons}: ($1\ to\ 10)\ KeV$
\par\noindent
{\rm Mean\ electron\ speed} $v/c \approx\ (0.06\ to\  0.2)$
\medskip
\par\noindent
Through the equipartition theorem, it was deduced that when the plasma contained deuterium ions, 
it had also a temperature equivalent to kinetic energies in the range $(1\div\ 10)\ KeV$, sufficient 
to overcome the Coulomb barrier and cause fusion. Reading from the graph of their integrated yield  
$Y$ of $2.5\  MeV$ neutrons, the yield seems to scale initially with the diameter $D$ according to 
$Y\propto\ 1/D^2$ and then it seems to saturate for higher diameters.  The authors state, without 
proof, that the variation in the yield  with fiber diameter can be understood on the basis of pressure 
and energetics. Their integrated yield $Y\sim\ 10^{10}$ neutrons of $2.5\  MeV$ for thin deuterated 
fibers of diameter $D\sim\ 20\ micron$. It is indeed a hefty rate giving ample proof that fusion had 
occurred. But they were puzzled by an anomalously high production of neutrons by natural i.e., non 
deuterated fibers as discussed below. 

\subsection{High neutron rates from non deuterated exploding wires \label{Nondeuterated}}

While the production of $2.5\ MeV$ neutrons from deuterated fibers appeared to satisfy expectations 
from the standard strong interaction branch Eq.(\ref{S1}), an apparent paradox  was generated when 
yields from normal fibers were found to be considerably higher than that expected. The isotopic 
abundance of deuterium in normal matter is about $1.5 \times\ 10^{-4}$. The fusion yield being 
proportional to the square of the density of deuterium, the expectations were that natural fiber yield 
should be about $5 \times\ 10^{-7}$ lower than the deuterated fiber yield, for the same fiber geometry. 
But measurements proved otherwise. The normal yields are about $4$ to  $5$ orders of magnitude higher than 
that expected. It is natural to conclude that some mechanism other than the strong interaction fusion is responsible 
for this effect. 

There are also other natural phenomena where neutrons have been observed in normal matter with virtually 
nil deuteron contamination. Examples are lightning\cite{Shah} and thunderstorms\cite{Gurevich}. In both 
cases, an anomalous production of neutrons has been reported.  For a normal fiber, if all we have are 
electrons and protons then the only way to produce neutrons is via weak interactions

\subsection{Normal Fibers and Weak Interactions \label{Normal}}

We shall discuss at some length the dynamics of weak neutron production in a later section. Here we 
shall just anticipate some results. As discussed earlier, for an electron-proton system, we are not 
afflicted by the Coulomb barrier, as a matter of fact the process is helped by Coulomb attraction. 
But to produce a neutron, about $0.78\ MeV$ needs to be put in the system to reach threshold for this reaction. 
To thermally excite an electron to such an energy is not possible even at the core of most 
stars. However, under special circumstances there are electromagnetic  means e.g., high electric and 
magnetic fields\cite{Pramana}, or ``smart'' materials\cite{rocky} which can directly convert elastic 
or other mechanical energy into EM signals, to supply such an energy.  

For a thin fiber, it is useful to consider an initial state of  a large number of ($N+1$) interacting electrons 
undergoing a weak process with a proton 
 \begin{equation}
\label{S2}
W + (N + 1) e^{-} + p \to\ N e^{-} + n + \bar{\nu}_e.
\end{equation}
We shall see later that the importance of having a large number of ``spectator'' electrons, even 
though only one is destroyed at a time, is due to the induction of a coherent Darwin 
interaction\cite{Darwin}  between the electrons which leads to a high collective contribution to 
the reaction energy thereby enhancing the nuclear activity. It produces a strong flux of neutrons 
in normal fibers, lightning and in thunderstorms.

\section {Electric LENT \label{ELENT}} 

In recent experiments, exploding molecular clusters of deuterium atoms have been produced. First, 
a weak laser pulse hits the cluster internally ionizing the atoms within the cluster and it is 
then followed by a strong laser pulse photo-ejecting a large number of electrons completely 
out of molecular cluster. This leaves the cluster with a large positive charge ($Q=\ Ne$) in a 
sphere of small radius $R$. Clusters with large $Q$ in a small $R$ explode. Ejected deuterons 
from different clusters then collide with sufficient kinetic energy to overcome the Coulomb 
barrier and lead to observed fusion events. Such experiments have now been repeated in several 
laboratories around the world and results bear out their theoretical analysis.

Let us estimate what magnitudes for the fields are required for a Coulomb explosion. Let $V$, $E$ and $P$ denote 
respectively the voltage, 
electric field and the stress on the system of charges.
 \begin{eqnarray}
\label{E1}
V = \frac{Q}{R} = \frac{Ne}{R}\ ,
\nonumber \\ 
E = \frac{Q}{R^2} = \frac{Ne}{R^2}\ ,
\nonumber \\
P = \frac{E^2}{8\pi} = \frac{N^2 e^2}{8 \pi R^4}\ .
\end{eqnarray}
The {\it tensile strength} of a material $P_c$ is defined as the maximum allowed stress before the material disintegrates.
\begin{eqnarray}
\label{E2}
E_c = \sqrt{8 \pi P_c},
\nonumber \\
for\ E<\ E_c,\ material \Rightarrow stable,
\nonumber \\
for\ E>\ E_c,\ material \Rightarrow unstable.
\end{eqnarray}
Typical explosion fields are of the order of
\begin{equation}
\label{E3}
E_c \sim\ 10^{11} Volts/m.
\end{equation}
We shall see later that similar fields would be required for inducing weak processes through surface plasmon polaritons on 
metallic hydride surfaces.

We can deduce a hot disintegration temperature  $T_c$ for Coulomb explosion devices, using Eqs.(\ref{E1}-\ref{E2}):
\begin{equation}
\label{E4}
eV_c = k_B T_c \sim \left(\frac{R}{100 \mathring{A}}\right) KeV \ \ \ \ T_c \sim\ 10^7 K,
\end{equation}
of the order of the temperature near the core of the Sun.

\subsection{Coulomb explosions of deuterium clusters \label{Coulomb}}

Through the collision between two deuterons from different clusters, Zweiback {\it et al} \cite{Zweiback} and 
Buersgens {\it et al} \cite{Buersgens}, 
observed the production of neutrons via 
the reaction Eq. (\ref{S1}):
\begin{equation}
\label{E5}
d + d \to\ ^3{\it He}(0.82 MeV) + n(2.45 MeV),
\end{equation}
the peak of their signal showing up at the expected characteristic neutron energy of $(2.45 \pm 0.2)\ MeV$. 

The cluster size can be measured through light scattering. When the wave length $\lambda$ of the incident light in a Raleigh light scattering 
experiment is large on the scale of the cluster radius $R$, then the elastic scattering amplitude is determined by the polarizability $\alpha$.
\begin{equation}
\label{E6}
F_{fi} (\omega) =\left (\frac{\omega}{c}\right)^2 \alpha(\omega) \epsilon^*_f \cdot \epsilon_i,
\end{equation}
so the elastic differential cross-section reads
\begin{equation}
\label{E7}
\left(\frac{d\sigma_{el}}{d\Omega}\right)_{i\to f} = |F_{fi} (\omega)|^2 =
\left (\frac{\omega}{c}\right)^4 |\alpha(\omega)|^2 |\epsilon^*_f \cdot \epsilon_i|^2
\end{equation}
and the integrated elastic cross-section reads
\begin{equation}
\label{E8}
\bar{\sigma}_{el} = \frac{8 \pi}{3}\left (\frac{\omega}{c}\right)^4 |\alpha(\omega)|^2 
\end{equation}
The total cross-section is given by
\begin{equation}
\label{E9}
\sigma_{tot} = \frac{4 \pi c}{\omega} \Im m\  F_{ii} = \frac{4\pi \omega }{c}  {\Im}m\ \alpha (\omega)
\end{equation}
The radius $R$ of the clusters has been determined empirically via light scattering cross sections as a function of the temperature $T$ of the gas 
jet producing the clusters. For a gas pressure of $55\ atmospheres$, as the temperature was increased from $85 K$ to $120 K$,  the measured 
radii of the clusters decreased from about $55 \mathring{A}$ to about $12 \mathring{A}$.

The temperature $T$ of the gas jet producing the clusters also determines the absorption cross section for the light. Thus, the cluster radius can 
be determined by the absorption of light. Such measurements have also been made. In addition, the yield as a function of the radius of the clusters 
have been measured. The experimental results are consistent with theoretical computations described above.

\subsection{Coulomb breakup of molecules \label{Molecules}}
The success achieved with deuteron fusion led to a series of experiments using other molecular clusters\cite{Esry,Nitrogen} through which 
detailed dynamical mechanisms for the explosions could be understood along with the structure of the molecules themselves. It has become an 
imaging tool as well, see for instance\cite{Legare}

\section{Electro-Weak LENT\label{EWLENT}}
A detailed review of electro-weak interaction induced lent [EW LENT] can be found\cite{Pramana}, 
hence we shall be very brief here. 
For systems containing only electrons and protons, as amply discussed in the preceding sections, 
weak interactions between electrons and protons are essential in producing neutrons. Once there 
is a supply of low energy neutrons, subsequent nuclear transmutations can occur at a high rate. 
It is crucial for the production of initial neutrons from low energy $e^-p^+$ in a condensed matter environment,
 that an input energy $W \geq 0.78\ MeV$ be supplied somehow to overcome the threshold 
barrier [see Eq.(\ref{G5})]. To accelerate the electrons thermally to MeV's in a condensed 
matter system is impracticable and one is left to search for EM means for the needed level of 
electron acceleration. There are essentially three options for electron acceleration which have 
already produced good results
\par\noindent
1. {\bf Fast Lasers}: Very fast femtosecond   lasers can and have been employed successfully to create mono-energetic electron beams from 
$100\ MeV$\cite{Nature1, Nature2, Nature3} up to $1\ GeV$\cite{Nature4} and table top electron accelerators have been constructed. While such 
table top accelerators are extremely useful devices for all sorts of studies, they are not quite suitable for inducing electroweak LENT. 
\par\noindent
2. {\bf Magnetic means}: As mentioned in the Introduction, an excellent example of EW LENT in nature is provided through the magnetic field 
acceleration of electrons and protons via the betatron mechanism  in the Solar corona. We shall discuss it in the next section (\ref{corona}).
\par\noindent
3. {\bf Electric means}: A laboratory example is an electric field acceleration of electrons when piezoelectric rocks are crushed. 
This would be discussed in Sec(\ref{Smart}) devoted to LENT induced by `smart' materials.

A very promising avenue for EW LENT yet not completely demonstrated in the laboratory is through plasmon polaritons on the surface
 of metallic hydrides. Under suitable conditions in particular, [$ecE/\Omega >\ \gamma_o (mc^2)$], where $E$ is the mean electric field, 
 $\Omega$ is the resonant frequency, $m$ is the electron mass and $\gamma_o\sim 2.5$, weak production of neutrons can occur. 
In fact, experimentally the Naples group has observed neutrons generated through plasma in a Mizuno type battery. The detection
of neutrons were made through nuclear CR-39 detectors and these results  are repeatable\cite{Cirillo2012}.  
Same as above  also holds for piezo electrics.The difference is that the resonant frequency for metallic hydrides is in the infrared whereas 
 acoustic frequencies are in the microwave range. Thus, considerably higher electric fields are needed for LENT to occur on metallic surfaces.   

\section{Solar Corona \label{corona}}  

\begin{figure}
\scalebox {0.6}{\includegraphics{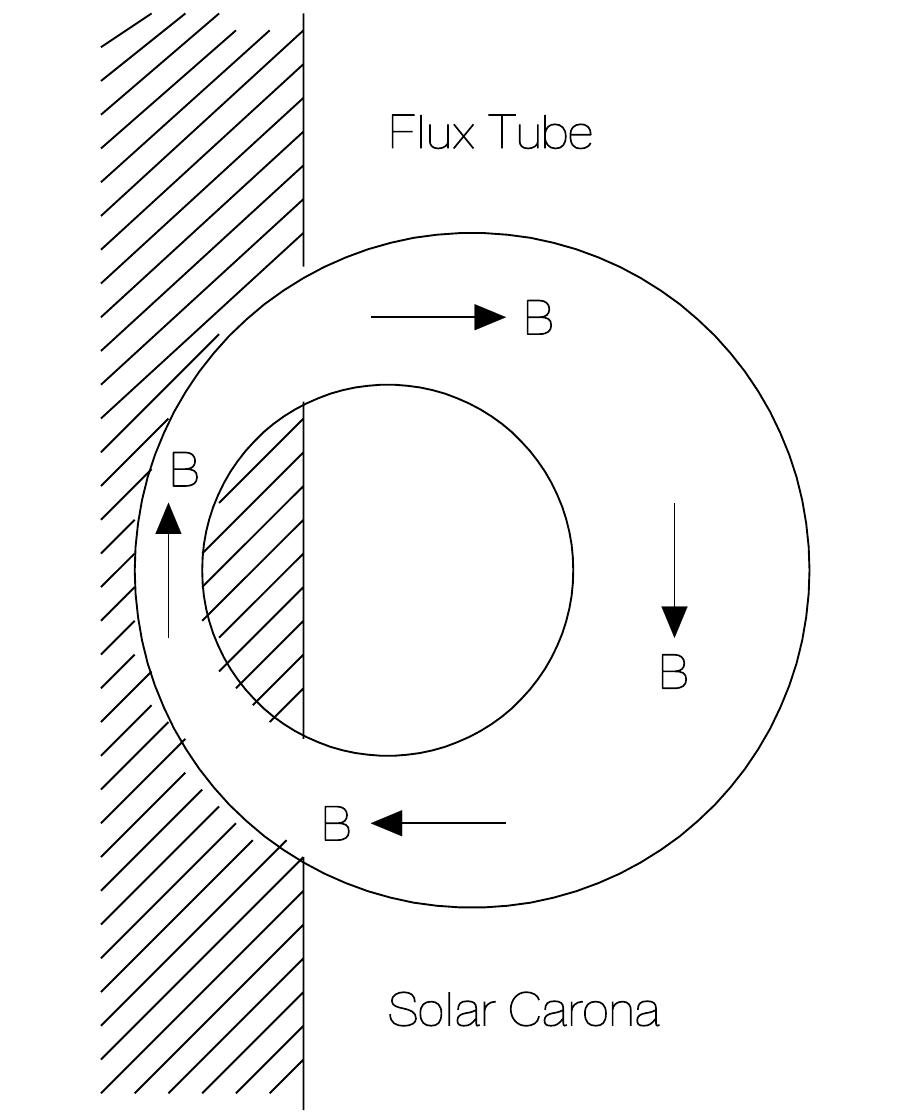}}
\caption{Shown schematically is a magnetic flux tube which
exits the solar photosphere (shaded region) and enters the
solar corona (clear region) through a sunspot on the chromosphere
boundary. The magnetic flux tube then exits the solar
corona and enters back into the photosphere through a second
sunspot on the chromosphere boundary. The walls of magnetic
flux tube are a vortex of circulating electric currents.} 
\label{fig2}
\end{figure}

Galileo first saw solar sunspots through his optical telescope and to this day the physical 
origin of these dark sunspots is unclear. Early in the last century, Hale and others\cite{Hale1, Hale2} 
discovered that magnetic flux tubes exit out of one sunspot to then dive into another sunspot. Such an 
uninterrupted magnetic flux tube has a part of it submerged below the solar surface where the 
electronic density is high and a part which is out in the solar corona where the density of electrons 
is low. A schematic picture is shown in FIG. \ref{fig2}. Such a ``floating'' flux tube is held up 
by the magnetic buoyancy\cite{Parker1}. There are oppositely directed vortex currents of electrons 
and protons which loop around the outer magnetic flux tubes. 

\subsection{Solar carpet \label{carpet}}

An intricate pattern of stable steady state magnetic flux tubes form a 
``magnetic carpet'' and they store considerable magnetic energy. This collective 
magnetic energy $W_{mag}$ can be estimated to be high enough to cause inverse 
beta decay similar to that discussed previously    
\begin{equation}
\label{s1}
W_{magnetic} + (N + 1) e^- + p \to  N e^- + n + \nu_e
\end{equation}
Let $\delta I$ denote a small change in the current going around the vortex circumference. 
It would lead to a small change in the 
magnetic field energy $\delta {\cal E}$
\begin{equation}
\label{s2}
\delta {\cal E} = \Phi\ \delta I.
\end{equation}
If $L$ denotes the length of the vortex circumference of the magnetic flux tube, then 
the destruction of one electron through the weak reaction Eq.(\ref{s1}) implies
\begin{equation}
\label{s3}
\delta I = - \frac{e v}{L},
\end{equation}
with $v$ denoting the component of the relative velocity between an electron and a 
proton which is tangential to the circumference. 
Letting $\Phi =\ B \Delta S $ and $\delta {\cal E} = - W_{magnetic}$, we have
\begin{equation}
\label{s4}
W_{magnetic} = (ec B) \left[\frac{\Delta S}{L}\right] \left[\frac{v}{c}\right].
\end{equation}
For a cylindrical flux tube, 
\begin{equation}
\label{s5}
\frac{\Delta S} {L} = \left[\frac{\pi R^2}{2 \pi R}\right] = \frac{R}{2}.
\end{equation}
yielding
\begin{eqnarray}
\label{s6}
W_{magnetic} = 15\ GeV \times 
\nonumber \\ 
\left[\frac{R}{Kilometer}\right] 
\left[\frac{B}{KiloGauss}\right] \left[\frac{v}{c}\right].
\end{eqnarray}
Employing the estimates 
\begin{equation}
\label{s7}
R \sim 100\ Km;\ B \sim\ 1\ KGauss;\ \frac{v}{c} \sim\ 10^{-2},
\end{equation}
we find
\begin{equation}
\label{s8}
W_{magnetic} \sim\ 15\ GeV
\end{equation}
that is quite large enough for an appreciable production of neutrons  from the protons. 
These neutrons thereby give rise to further nuclear transmutations into higher mass nuclei 
via neutron capture and subsequent beta decays.

\subsection{Betatron mechanism \label{Betatron}}

Let us discuss here a non-thermal acceleration process which leads to a generation of quite 
energetic particles in the solar corona. As described in the last section, a mechanism is 
required to induce nuclear reactions well outside the solar photosphere. 

The central feature of our solar accelerator \cite{WSL}  centers around the Faraday's 
law and is closely analogous to a betatron\cite{Kirst}. Conceptually, the betatron is a 
step up transformer whose secondary coil is a toroidal ring of accelerating charged particles 
circulating about a time varying magnetic flux tube. The solar magnetic flux tube acts 
as a step up transformer mechanism in transferring particle kinetic energy outward from the 
inner photosphere to circulating charged particles located in the corona outside.
 
Circulating  currents located deep in the photosphere can be viewed conceptually as a net 
current $I_{primary}$ circulating around a primary coil. Circulating currents found high in 
the corona can be viewed as a net current 
$I_{secondary}$ circulating around a secondary coil. If $K_{primary}$ and $K_{secondary}$ represent,respectively, 
the charged particle kinetic energies in the primary and secondary coils, 
then one finds the step up transformer power equation 
\begin{eqnarray}
\label{B1}
{\dot K}_{primary} = I_{primary} V_{primary}\nonumber\\ 
= {\dot K}_{secondary} = I_{secondary} V_{secondary}
\end{eqnarray}
where $V_{primary,secondary}$ represent the voltages across the primary, secondary coils 
respectively. The total kinetic energy transfer
\begin{eqnarray}
\label{B2}
\Delta K_{primary} = \int I_{primary} V_{primary} (dt)\nonumber\\
 =  \Delta K_{secondary} = \int I_{secondary} V_{secondary} (dt),
\end{eqnarray}
The essence of the step up transformer mechanism is that the kinetic energy distributed among a 
very large number of charged particles in the photosphere can be transferred via the magnetic 
flux tube to a distributed kinetic energy shared among a distant much smaller number of charged 
particles located in the corona, i.e. a small accelerating voltage in the primary coil produces 
a large accelerating voltage in the secondary coil. The resulting transfer of kinetic energy is 
collective from a large group of charged particles to a smaller group of charged particles. The 
kinetic energy per charged particle of the dilute gas in the corona may then become much higher 
than the kinetic energy per charged particle of the more dense fluid in the photosphere. In terms 
of the connection between temperature and kinetic energy, the temperature of the dilute
gas in the corona will be much higher than the temperature of the more dense fluid photosphere.

It is clear that for the betatron, or equivalently the transformer mechanism, to be efficient in 
sustaining large electrical fields, the coronal electrical conductivity must not be too large. For 
an estimation of the conductivity, we need to know the density of the electrons in the corona. 
Fortunately, we have a quite reliable experimental measurement  of the electronic
density through its effect on the gravitational bending of light in the radio wave band near the 
Sun. Given its importance, we shall sketch the methodology of this measurement. 

Bending of an EM beam near the Sun may be looked upon as a change in the refractive index, from 
unity for the vacuum. For frequencies $\omega>>\ \omega_P$, where $\omega_P$ is the plasma frequency, 
the effective refractive index
${\tilde n}({\bf r}, \omega)$ at a position ${\bf r}$, where the charged particle density is 
$n({\bf r})$,  reads \cite{Weinberg, Muhleman}
\begin{equation}
\label{B3}
{\tilde n}({\bf r}, \omega) = 
1 + \frac{2 G M_{Sun}}{c^2 |{\bf r}|} - \frac{2\pi n({\bf r})e^2}{m \omega ^2}.
\end{equation}
Successful measurements of the gravitational bending of electromagnetic waves with frequencies in 
the visible and all the way down to the high end of the radio spectrum are legion. These experiments 
provide a direct proof that any coronal conductivity disturbance on the expected gravitational bending 
of electromagnetic waves for frequencies down to $12.5$ GigaHertz must be negligible. Deviations from 
the general relativistic effect have allowed quite accurate measurements of the electronic density 
in the Solar corona.  

The estimates from even lower frequency radio wave probes used for gravitational bending \cite{Muhleman} 
put the coronal conductivity in the megahertz range. For comparison, we note that the typical conductivity 
of a good metal would be more than ten orders of magnitude higher\cite{Salt}. Hence, these data lead us to 
conclude that the Solar corona is close to being an insulator and eons away from being a metal and there 
are no impedimenta toward sustaining electrical fields within it \cite{Foukal, Feldman, Stasiewicz}. Thus, 
our proposed transformer mechanism and its subsequent predictions for the corona remain intact.

\subsection{Solar flares and Solar explosions \label{flares}}

When sufficiently high kinetic energy is reached in the circulating currents around the flux tubes 
floating in the corona, the flux tubes may explode violently into a solar flare. The loss of magnetic 
energy during the flux tube explosion is rapidly converted into charged particle kinetic energy. The relativistic 
high energy products of the explosion yield both nuclear and elementary particle 
interactions. The extraordinary explosion which occurred on July 4, 2000 was strong enough to 
have generated particles, for example an excess muon flux, some of which were observed experimentally 
at the CERN L3 + C detector\cite{L3}. It offered a wonderful opportunity to infer that protons with an 
average energy of $40\ GeV$ were produced in the Solar corona. Independently, the BAKSAN underground 
muon measurement\cite{Baksan} provided experimental evidence for protons of energy in excess of 
$500\ GeV$ in the Solar flare of September 29, 1989. We show below some estimates\cite{Pramana}.

For a large Solar flare exploding in a time ($\Delta t$), the break up of a magnetic tube would yield, 
through the Faraday law, a mean acceleration voltage ${\bar V}$ around the walls given by
\begin{equation}
\label{So1}
{\bar V} = \frac{\Delta \Phi}{\Delta t}
\end{equation}
Inserting as before, $\Delta \Phi\ = B \Delta S$, where $B$ denotes the mean magnetic field before 
the explosion and $\Delta S$ the inner cross-sectional area, we have for the mean acceleration energy
\begin{equation}
\label{So2}
e {\bar V} = ecB\ \frac{\Delta S}{\Lambda}\ \ \ {\rm where} \ \ \ \Lambda = c \Delta t .
\end{equation}
For a cylindrical geometry, we may again rewrite it in a useful form 
\begin{equation}
\label{So3}
e {\bar V} \approx 30\ GeV\times \left[\frac{B}{KiloGauss}\right] 
\left[\frac{\pi R^2}{\Lambda \times Kilometer}\right].  
\end{equation} 
For a coronal mass ejecting flux tube exploding in a time 
$\Delta t\ =\ 10^2$ seconds, we use the estimated values to derive
\begin{eqnarray}
\label{So4}
B \approx\ 1\ KiloGauss;\nonumber\\
R \approx\ 10^4\ Kilometers\nonumber\\
\Lambda \approx\ 3 \times\ 10^7\ Kilometers\nonumber\\
e {\bar V} \approx\  300\ GeV 
\end{eqnarray}
Physically, it corresponds to a colliding beam of electrons and protons with a beam 
energy of $300\ GeV$. Experimental measurements by L3 and Baksan confirm the very high energy 
acceleration of electrons and protons and our proposed mechanism is consistent with the data. 
In \cite{Pramana} we have also made predictions for the flux of the positron from such flares. 

Needless to say, that the measurements already made belie the notion that all nuclear reactions 
in the Sun occur in the core of the Sun. Similar conclusions can be drawn for any star with a 
high enough magnetic field.

\subsection{Fusors}

It is often overlooked that in a D-D or a D-T 
system\cite{IEEE-D-3He} -driven by modest static electric fields-
inertial electrostatic confinement has been achieved leading to fusion. Such devices known as
 fusors have existed since the 1960's. The basic idea is to use a hollow transparent (perforated)
spherical cathode enclosed by a concentric spherical anode, and the device
filled with deuterium or deuterium and tritium at low pressure. Ionization of
the gas occurs via corona discharge. A potential
difference of around 80 KV in commercial devices  accelerates
deuterons radially, and despite collisional losses, 
substantial number of collisions take place with energies over 15 KeV and fusion takes place 
with the release of neutrons. Much lower voltages can be used. 
In fact, the voltage threshold for D-T fusion (the lowest) is 
very modest, comparable to that used
in many photocopiers, well below the 10 KV of old black and white
television sets or furnace ignition transformers, and also well below the
typical 15 KV for neon signs or the 25 KV used in old-style colour TV sets. 
For D-D fusion, the threshold voltage is again quite modest around 20 KV, 
and corresponds to an energy that can and indeed has been reached by pyroelectric devices 
as discussed in section \ref{Pyro}. 

We stress that this is not a speculative idea, but rather a mature technology, 
and tabletop devices can be purchased commercially\cite{NSD-Fusion} 
which produce ($10^6\div 10^{11}$) neutrons per second. Indeed, 
a quick internet search will readily reveal that fusors are popular science
fair projects built by students.

\section{LENT in Smart Materials\label{Smart}}

Amongst a large class of ``smart'' or more appropriately, specifically designed materials with 
special properties, pyro and piezo electric materials have been demonstrated both theoretically 
as well as experimentally to lead to low energy nuclear reactions. A pyroelectric crystal develops 
an electric field due to (adiabatic) changes in its temperature and its opposite, that is an 
applied electric field causing an adiabatic heating or cooling of the system is called the 
electrocaloric effect. In a piezo electric crystal, mechanical stress is directly converted to the 
development of electric fields and vice versa. Under suitable conditions, sufficient electric fields [voltages] have 
been generated in both cases to lead to neutron production. We shall discuss each 
of them briefly.  

\subsection{Pyroelectric LENT \label{Pyro}}

In \cite{Naranjo, Geuther}, it was experimentally shown that through two pyroelectric crystals when 
heated or cooled produced nuclear $d d$ fusion evidenced by the observation of the characteristic 
signal of $2.5\ MeV$ neutrons. The system was used to ionize the gas and accelerate the ions up 
to $200\ KeV$. As we have seen in the previous sections, e.g.  Eq.(\ref{S1}), this acceleration is 
much more than sufficient to cause $d d$ fusion. The measured yields are in good agreement with the 
calculated yield.

The formal definition of a pyroelectric crystal is that within a single domain, the electric 
induction $<{\bf D}_o> \ \neq 0$. When such a pyroelectric crystal is heated or cooled, it gets 
spontaneously polarized under equilibrium conditions. The often quoted phenomenological expression 
presumably valid in the linear regime gives the effective electric field $E_{eff}$ generated inside 
a pyroelectric crystal due to a temperature change $\Delta T$ 
\begin{equation}
\label{P1}
E_{eff} = \phi (\Delta T),
\end{equation}
where $\phi$ is the pyroelectric coefficient. Pyroelectric crystals are modeled as parallel plate 
capacitors with the accelerating voltage $V$ given by
\begin{equation}
\label{P2}
V = \frac{Q}{C}\ ,\ \ \ \ Q = E_{eff} A, \ \ \ \ C = \frac{A \epsilon}{4 \pi t}\ , 
\end{equation}
where $A$ is the area, $\epsilon$ is the dielectric constant and $t$ the thickness of the crystal. 
Hence, in this simple model, the energy imparted to an electron or an ion of charge $|e|$ reads 
\begin{equation}
\label{P3}
|e| V = \frac{4 \pi e \phi t (\Delta T)}{\epsilon} 
\end{equation}
If two crystals are used face to face, then the energy above is doubled.

For definiteness, we shall discuss briefly the experiment described in\cite{Geuther}, where lithium 
tantalate [$LiTaO_3$] crystals were employed since they have a particularly large value for this 
coefficient,  
\begin{eqnarray}
\label{P4}
\phi \approx 190\ \frac{\mu C}{m^2\ K} \approx 57\ \frac{Gauss}{K}
\nonumber\\ 
\phi \approx 1.71 \times 10^4\  \frac{Volts}{cm\ K}.
\end{eqnarray}
At atmospheric pressure, the change in polarization charge is quickly masked by the accumulation 
of charges from the air. Hence, it is in the vacuum that we see the surface charge change and the 
resulting electric field. 
 
For, $t = 1\ cm$ and $\epsilon = 46$ for a two crystal set up\cite{Geuther}, the increase in 
energy should be
\begin{equation}
\label{P5}
2 |e| V = 933\ KeV,
\end{equation} 
whereas the maximum electron energy gain measured is only about $200\ KeV$. For ions, it is even 
less.

The deuterium fill gas in a vacuum chamber  was ionized by the strong electric field between two 
polarized crystals and the deuterons were accelerated into a deuterated polystyrene target at 
energies above the threshold for fusion. To obtain the maximum accelerating voltage, two crystals 
were employed to double the voltage and  a catwhisker tip was employed to increase the electric 
field. The crystals were heated up to $130 ^o C$ and then cooled to room temperature. Very clear 
signals for neutrons with energies $(2.5 \pm 0.1)\ MeV$ were observed by the usual time of flight 
method.

\subsection{Piezoelectric LENT \label{Piezo}}

Over six decades, considerable experimental evidence has been gathered for high energy particle 
production during the fracture of certain kinds of crystals 
\cite{Karassey, Klyuev, Preparata, Nakayama, Lawn}. 
Fracture induced nuclear transmutations and the production of neutrons have been clearly 
observed\cite{Derjaguin, Lipson, Kaushik, Shioe, Cardone, Fuji}. The production of neutrons 
appears greatly enhanced if the solids being fractured are piezoelectric materials. Common 
examples of a piezoelctric are : hair, bone, quartz. 

In a recent paper\cite{rocky}, we have theoretically analyzed the problem and shown the manner 
in which the mechanical pressure in a piezoelectric stressed solid about to fracture can organize 
the energy so that neutrons can be produced. Since some of the arguments may not be familiar to 
many readers, we shall repeat below some of the relevant theory. 

The energy per unit volume of a piezoelectric material may be written as
\begin{equation}
\label{p1}
du =  T ds + \sigma_{ij} dw_{ij} - {\bf P}{\cdot}  d{\bf E},
\end{equation}
where $s$ is the entropy per unit volume;  $\sigma_{ij}$ and $w_{ij}$ are the stress and strain 
tensors, ${\bf P}$ and ${\bf E}$ are the polarization and electric field vectors. The adiabatic 
piezoelectric tensor is defined as\cite{LL}
\begin{equation}
\label{p2}
\beta_{i, jk} = 
\left(\frac{\partial P_i}{\partial w_{jk}}\right)_{s, {\bf E}} 
= - \left(\frac{\partial \sigma_{jk}}{\partial E_i}\right)_{s, w} 
\end{equation}
The effective mechanical electric field interaction energy density, up to quadratic order, 
reads
\begin{eqnarray}
\label{p3}
\beta_{i, jk} = - \left(\frac{\partial^2 u}{\partial E_i \partial w_{jk}}\right) 
= - \left(\frac{\partial^2 u}{\partial w_{jk} \partial E_i}\right), 
\\
u = - \beta_{i, jk} E_i w_{jk},
\end{eqnarray}
leading to the piezoelectric Hamiltonain 
\begin{equation}
\label{p4}
{\cal H}_{piezo} = - \int (d^3{\bf x}) \beta_{i, jk} E_i w_{jk}.
\end{equation}
Eq.(\ref{p4}) provides a direct link between a large strain (say a fracture caused by stress) 
causing a large electric field. If one uses this interaction twice, then to second order in 
$\beta$, the EM photon propagator would receive a contribution from the mechanical acoustic 
phonon propagator in the following way.

Let $\chi_{ij}$ and ${\tilde \chi}_{ij}$ be the adiabatic electric susceptibilities at 
constant strain and at constant stress respectively
\begin{equation}
\label{p5}
\chi_{ij} = \left(\frac{\partial P_i}{\partial E_j}\right)_{s,w},
\ \ \ \ \ \ \  
{\tilde \chi}_{ij} = \left(\frac{\partial P_i}{\partial E_j}\right)_{s,\sigma}.
\end{equation}
These two are related through the elastic response tensor
\begin{equation}
\label{p6}
D_{ijkl} = \left(\frac{\partial w_{ij}}{\partial \sigma_{kl}}\right)_{s,{\bf E}} \ ,
\end{equation}
via the thermodynamic identity
\begin{equation}
\label{p7}
{\tilde \chi}_{ij} = \chi_{ij}  + \beta_{i,lm} D_{lm, nq} \beta_{j, nq}.
\end{equation}
The above static equations are easily generalized to dynamical frequency dependent 
susceptibilities $\chi_{ij}(\omega)$ and ${\tilde \chi}_{ij}(\omega)$ and the phonon 
propagator $D_{lm, nq} (\omega)$. Phonon modes described by the phonon propagator 
affect the dynamical susceptibilities through the relations 
\begin{eqnarray}
\label{p8}
{\bf D} =  {\bf E} + 4 \pi {\bf P}
\\
\epsilon_{ij} (\omega) = \delta_{ij} + 4 \pi {\tilde \chi}_{ij}(\omega)
\\
{\tilde \chi}_{ij}(\omega) = \chi_{ij}(\omega)  
+ \beta_{i,lm} D_{lm, nq}(\omega) \beta_{j, nq}.
\end{eqnarray}
Clearly then, the electric susceptibility of a piezoelectric material would be affected 
by the frequencies associated with the mechanical acoustic frequencies occurring in 
fractures of such materials. The velocity of sound $v_s$ compared to the 
velocity of light $c$ obeys $v_s/c\approx\ 10^{-5}$ and hence for similar sized cavities
\begin{equation}
\label{p9}
\left(\frac{\omega_{phonon}}{\omega_{photon}}\right) \sim 10^{-5}
{\rm \ for\ similar\ sized\ cavities}.
\end{equation}
These phonon modes will then appear in $\epsilon(\omega)$.

The nuclear reactions are supposed to proceed through the weak capture of high electric 
field accelerated electrons by protons as described in Section(\ref{EWLENT}). We need to 
calculate the mean energy of the electrons in condensed matter when accelerated by an 
electric field
\begin{equation}
\label{p10}
\frac{d {\bf p} }{dt} = e {\bf E},
\end{equation}
and the electron energy is estimated to be
\begin{equation}
\label{p11}
W = \sqrt{(mc^2)^2 + < |{\bf p}|^2> c^2}.
\end{equation}
If ${\cal P}_E(\omega) (d\omega)$ denotes the mean squared electric field strength in a 
band width ($d\omega$), then Eqs.(\ref{p10},\ref{p11}) imply
\begin{eqnarray}
\label{p12}
E^2 = \int_o^\infty d\omega {\cal P}_E(\omega)
\\
< |{\bf p}|^2> = e^2 \int_o^\infty \frac{d\omega}{\omega^2} {\cal P}_E(\omega).
\end{eqnarray}
If $\Omega$ is a dominant frequency, then the above equation may be written as
\begin{equation}
\label{p13}
 < |{\bf p}|^2>  = \frac{e^2 E^2}{\Omega^2}.
\end{equation}
Let $m^*$ be the (electrically) renormalized mass. Then, in terms of the vacuum electron 
mass $m$, we may write 
\begin{equation}
\label{p14}
m^* = \gamma\ m\ \ \ {\rm where} 
\ \ \gamma = \sqrt{1 + \left(\frac{e E}{mc \Omega}\right)^2}
\end{equation}
For electron capture to occur, $\gamma \geq \gamma_o \sim\ 2.5$. For a rough estimate of the 
electric field, we may assume that most of the stress at a fracture is electric, in which case
\begin{equation}
\label{p15}
\sigma_{fracture} \sim \frac{E^2}{8 \pi}\ \ \ \Rightarrow 
\ \ \ E \sim 10^5\ Gauss, 
\end{equation}
where we have used Griffith's rule that 
$\sigma_{fracture} \sim 10^{-3}\ \sigma_{bond}$, where $\sigma_{bond}$ is the stress 
necessary to break all the chemical bonds. For a relevant frequency in the microwave 
range $\Omega \sim 6 \times\ 10^{10}/sec$ we have $\gamma \sim 30$. We can convert this 
into the neutron production rate through the Fermi interaction
\begin{eqnarray}
\label{p16}
\Gamma [{\tilde e} + p \to n + \nu_e] 
\sim\ \left[\frac{G_F m^2}{\hbar c}\right]^2 
\left[\frac{mc^2}{\hbar}\right] \gamma^2\\
\sim 0.6\ Hz\ {\rm for}\  \gamma = 30.
\end{eqnarray}
The transition rate per unit time per unit area of micro-crack surface may be written as
\begin{equation}
\label{p17}
\nu_2 = n_2 \Gamma [{\tilde e} + p \to\ n + \nu_e] ,
\end{equation}
where $n_2$ is the number of protons per unit area in the first few layers of the quartz granite. 
Typical values are
\begin{equation}
\label{p18}
n_2 \sim \frac{10^{14}}{cm^2}\ \ \ \Rightarrow 
\ \ \ \nu_2 \sim \frac{10^{15} Hz}{cm^2} \ , 
\end{equation}
a neutron production rate sufficiently large to be measured and to cause even other nuclear transformations. 
Incidentally, in the piezoelectric case, the neutrons are not ultra cold.

\subsection{Novel uses for the Tokamak \label{Tokomak}}

Traditional ``hot fusion'' programs have been able to generate neutrons at total rate of about 
$10^{17}/sec.$ in the ($dd$) and ($dt$) modes, unfortunately lasting only for about 2 seconds
\cite{Fusion1, Fusion2}. Thus, great improvements are still required for a steady state running 
of the Tokamaks along with the curtailment of the generated radioactivity. As 
emphasized by Byrne\cite{Byrne}, fusion reactors would require a ``biological shield'',
as their outermost shell made of some heavy element such as lead, whose purpose is to remove 
the surviving $14\ MeV$ neutrons and to attenuate the $\gamma$ ray flux. But, this would generate 
a substantial amount of undesirable radioactivity\cite{Steiner}. In fact, Byrne concludes that 
``{\it It is not immediately obvious that the problems of radioactive waste disposal, which have 
plagued fission power programmes across the world, will be entirely absent in the thermonuclear power 
plants of the future.} '' 
 
In view of the formidable scientific and technical problems with hot fusion programs yet to be 
resolved, it is pertinent to ask whether Tokomaks, machines constructed at great public cost, may 
be put to a different but important practical use. Given the extremely large electromagnetic fields 
created with great ingenuity within these devices, we propose to use it as a Proof of Concept machine 
for electro-weak LENT by running electrons and protons rather than just deuterons and tritons as has 
been done so far. It may be difficult at the beginning to adapt a Tokomak for this purpose 
but our proposal seems feasible.  

\section{Conclusions and Future Prospects \label{Conclusion}}
The leitmotif of this paper has been to provide direct evidence that low energy nuclear 
transmutations [LENT] can and have been induced by gravitational, weak, EM and strong 
interactions both in Nature as well as in the laboratory. Also, that quite often three 
or four of the fundamental interactions are needed to lead to the final LENT. 
Due to this interplay of different interactions each with its own widely different coupling 
strength, one finds a great diversity in the rates of the reactions.

Hence, we may conclude that the reality of LENT is no longer a matter of debate because 
it is soundly based on the Standard Model.The focus must now  be on the employment of 
technology to realize electroweak LENT and thereby provide sorely needed practical applications 
of the Standard Model of particle physics. Extraordinary advances already made in realizing up to 
$1\ GeV$ table top electron accelerators and high gain factors in generating electric fields in 
surface plasmon polariton resonances on nano particles bode well for the future of electroweak LENT.

\section{Acknowledgements}

YS would like to thank his colleagues at the U. of Rome La Sapienza, Pisa, Perugia, Naples, Boston 
and at CERN for inviting him for seminars on this and related topics. Special thanks are due in 
particular to Professor N. Cabibbo, who during his last years greatly encouraged YS to pursue 
ideas presented here and to, Dr. R. Chidambaram who has been an ever patient and sympathetic listener.


\begin{thebibliography}{11}



\bibitem{Bethe} 
H. A. Bethe and C. L. Critchfield, {\it Phys. Rev.}, {\bf 54}, 248 (1938); 
H. A. Bethe, {\it Phys. Rev.}, {\bf 55}, 436 (1939). 

\bibitem{BetheNobel} 
H. A. Bethe, {\it Energy Production in Stars}, Nobel Lecture, December 11, 1967.

\bibitem{Stephanakis} 
S. J. Stephanakis, L. S. Levine, D. Mosher, I. M. Vitkovitsky, 
and F. Young, {\it Phys. Rev. Lett.}, {\bf 29}, 568 (1972).

\bibitem{Pramana} 
Y. N. Srivastava, A. Widom, L. Larsen, {\it Pramana} {\bf 75}, 617 (2010); 
arXiv: nucl-th 0810.0159.

\bibitem{Zweiback} 
J. Zweiback, R. A. Smith, T. E. Cowan, G. Hays, K. B. Wharton, V. P. Yanovsky 
and T. Ditmire, {\it Phys. Rev. Lett.}, {\bf 84}, 2634 (2000); 
{\it ibid} {\bf 85}, 3640 (2000).


\bibitem{WSL} 
A. Widom, Y. N. Srivastava and L. Larsen, 
{\it High Energy Particles in the Solar Corona}, arXiv:nucl-th 0804.2647. 

\bibitem{Wurz} 
P. Wurz {\it et al}, {\it Geophysical Research Letters} {\bf 25}, 2557 (1998).

\bibitem{Cowley} 
C. Cowley, W. Bidelman, S. Hubrig, G. Mathys, and D.Bord, 
{\it Astronomy and Astrophysics} {\bf 419}, 1087 (2004).

\bibitem{Goriely} 
S. Goriely, {\it Astronomy and Astrophysics} {\bf 466}, 619 (2007).

\bibitem{Lodders} 
K. Lodders, arXiv:0710.4523 (October 2007).


\bibitem{L3} 
L3 + C Collaboration, {\it Astronomy and Astrophysics} {\bf 456}, 357 (2006).

\bibitem{Kelvin} Lord Kelvin [W .Thomson], ``1861 Report of the 31st Meeting of the British 
Association for the Advancement of Science'', John Murray, London; Part II, page 27.

\bibitem{AgeSun} 
{\it Astrophysical Ages and Times Scales}, ASP Conference Series {\bf 245}, 
Edited by Ted von Hippel, Chris Simpson, and Nadine Manset, 
San Francisco: Astronomical Society of the Pacific, ISBN: 1-58381-083-8, 31 (2001).

\bibitem{Alvarez} 
L. Alvarez, {\it Phys. Rev.} {\bf 54}, 486 (1938).


\bibitem{Weizsacker}  
C. F. von Weizs\"acker {\it Zeit. f$\ddot{u}$r  Physik.}, {\bf 38} 176 (1937).

\bibitem{Fowler} 
W. A. Fowler, G. R. Caughlan and B. A. Zimmerman, 
{\it Ann. Rev. Astron. Astrophys.} {\bf 5} 525 (1967).

\bibitem{WL} 
A. Widom and L. Larsen, {\it Eur. Phy. J.} {bf C46}, 107 (2006).  
\ arXiv: cond-mat/050502; A. Widom and L. Larsen, arXiv: nucl-th/0608059v2.

\bibitem{Schwinger} 
J. Schwinger, in Fourth International Conference 
ICCF4, Maui, Hawaii, December 1994.  

\bibitem{Bahcall} J. Bahcall, {\it Phys. Rev.}, {\bf 128} 1297 (1962).

\bibitem{Selove} 
F. Ajzenberg-Selove and T. Lauritsen, {\it Nuc. Phys.} {\bf 11}, 1 (1959).

\bibitem{Weizsacker1} 
C. F. von Weizs\"acker, {\it Zeit. f$\ddot{u}$r  Physik.} {\bf 39}, 633 (1938).

\bibitem{Bethe1} 
H. A. Bethe, {\it Phys. Rev.} {\bf 55}, 436 (1939).

\bibitem{Salpeter} 
E. E. Salpeter, {\it Phys. Rev.} {\bf 88}, 547 (1952).

\bibitem{Hoyle} 
F. Hoyle, {\it Astrophys. J., Suppl.} {\bf 1}, 121 (1954).


\bibitem{Shah} 
G. N. Shah, H. Razdan, C. L. Bhat and Q. M. Ali, {\it Nature, (London)} {\bf 313}, 773  (1985).

\bibitem{Gurevich} 
A. V. Gurevich, V. P. Antonova, A. P. Chubenko, A, N. Karashtin, G. G. Mitko, 
M. O. Ptitsyn, V. A. Ryabov, A. L. Sheptov, Yu. V. Shlyugaev, L. I. Vildanova, 
and K. B. Zybin, {\it Phys. Rev. Lett.} {\bf 108}, 125001 (2012).

\bibitem{rocky}  
A. Widom, J. Swain and Y. N. Srivastava, {\it Neutron Production from the Fracture 
of Piezoelectric Rocks}, arXiv:1109.4911v2 [physics.gen-ph]

\bibitem{Darwin} 
C. D. Darwin, {\it Phil. Mag.} {\bf 39}, 537 (1920). 

\bibitem{Buersgens} 
F. Buersgens, K. W. Madison, D. R. Symes, R. Hartke, J. Osterhoff, W. Grigsby, 
G. Dyer and T. Ditmire , {\it Phys. Rev. E} {\bf 74}, 016403  (2006).
 
\bibitem{Esry}  
B. D. Esry {\it et al} {\it Phys Rev Lett} {\bf 97}, 013003 (2006).
 
\bibitem{Nitrogen} 
Wu C. {\it et al}, {\it Phys Chem Chem Phys} {\bf 13(41)}, 18398 (2011).
 
\bibitem{Legare} 
F. Leg'are {\it et al}, {\it Phys Rev} {\bf A71}, 013415 (2005).
 
\bibitem{Nature1} 
S. Mangles {\it et al}, {\it  Nature} {\bf 431}, 535 (2004).
 
\bibitem{Nature2} 
C. Geddes {\it et al}, {\it  Nature}, {\bf 431}, 538 (2004).

\bibitem{Nature3} 
J. Faure {\it et al}, {\it  Nature} {\bf 431}, 541 (2004).

\bibitem{Nature4} 
W. Leemans {\it et al}, {\it Nature Physics} {\bf 2}, 696 (2006).

\bibitem{Cirillo2012}
C. Domenico, R. Germano, V. Tontodonato, A. Widom, Y. N. Srivastava, E. Del Giudice, G. Vitiello
{\it Key Engineering Materials} {\bf 495}, 104 (2012).

\bibitem{Hale1} 
G. E. Hale, {\it Astrophys. J} {\bf 28}, 315 (1908).

\bibitem{Hale2} 
G. E. Hale, F. Ellerman, S. B. Nicholson and A. H. Joy, {\it Astrophys. J} {\bf 49}, 153 (1919).

\bibitem{Parker1} 
E. N. Parker, {\it Astrophys. J} {\bf 121}, 491 (1955).

\bibitem{Kirst} 
D. W. Kirst, {\it Phys. Rev.} {\bf 60}, 47 (1941);  D. W. Kirst and R. Serber, 
{\it Phys. Rev.} {\bf 60}, 53(1941).

\bibitem{Weinberg} 
S. Weinberg, {\it Gravitation and Cosmology: Principles and Applications of the 
General Theory of Relativity}, Chapter 8. Published by John Wiley and Sons, New York (1972).

\bibitem{Muhleman} 
D. O. Muhleman, R. D. Ekers and E. B. Fomalont, {\it Phys. Rev. Lett.} {\bf 24}, 1377 (1970).

\bibitem{Salt} 
By way of comparison, salty ocean water has a conductivity of about $1.5 \times\ 10^{11}\ Hertz$.

\bibitem{Foukal} 
P. Foukal and P. Miller, {\it Solar Physics} {\bf83}, 83 (1983).

\bibitem{Feldman} 
U. Feldman, {\it Physica Scripta} {\bf 65}, 185 (2002).

\bibitem{Stasiewicz} 
K. Stasiewicz and J. Ekeberg, {\it Astrophys. J.} {\bf 680}, L153 (2008).

\bibitem{Baksan} 
S. N. Karpov {\it et al}, {\it Il Nuovo Cimento} {\bf 21C}. 551 (1998).

\bibitem{IEEE-D-3He} See, for example: R. P. Ashley, G. L. Kulcinski, J. F. Santarius, S. Krupakar Murali, G. Piefer; IEEE Publication 99CH37050, pg. 35-37, 18th Symposium on Fusion Engineering, Albuquerque NM, 25-29 October 1999.

\bibitem{NSD-Fusion} {\tt http://www.nsd-fusion.com/core-tech.php}

\bibitem{Naranjo} 
B. Naranjo, J. Gimzewski and S. Putterman,  {\it Nature (London)} {\bf 434}, 1115 (2005).

\bibitem{Geuther} 
J. Geuther, Y. Danon, and Frank Saglime,  {\it Phys. Rev. Lett.} {\bf 96}, 054803 (2006).

\bibitem{Karassey} 
V. V. Karassey, N. A. Krotova and B. W. Deryagin, 
{\it Dokl. Akad. Nauk SSSR} {\bf 88}, 777 (1953).

\bibitem{Klyuev} 
V. A. Klyuev, A. G. Lipson, Yu. P. Toporov, B. V. Deryagin, V. J. Lushchikov, 
A.V. Streikov, E. P. Shabalin, {\it Sov. Tech. Phys. Lett.} {\bf 12}, 551 (1986); 
V. Klyuev {\it et al.}, {\it Kolloidn. Zh.} {\bf 88}, 1001 (1987).

\bibitem{Preparata} 
G. Preparata, {\it Il Nuovo Cimento}, {\bf 104}, 1259 (1991).

\bibitem{Nakayama} 
K. Nakayama, N. Suzuki and H. Hashimoto, {\it Journal of Physics D} {\bf 25}, 303 (1992).

\bibitem{Lawn} 
B. Lawn, {\it Fracture of Brittle Solids}, Sec. 4.5, page 103, 
Cambridge University Press, Cambridge (1993).

\bibitem{Derjaguin} 
B. V. Derjaguin, A. G. Lipson, V. A. Kluev, D. M. Sakov and Yu. P. Toporov, 
{\it Nature} {\bf 341}, 492 (1989).

\bibitem{Lipson} 
A. G. Lipson, D. M. Sakov, V. A. Klyuev and B. V. Deryagin, 
{\it JETP. Lett.} {\bf 49}, 675 (1989); B. V. Derjaguin, V. A. Kluev, 
A. G. Lipson, and Yu. P Toporov, {\it Physica B} {\bf 167}, 189 (1990).

\bibitem{Kaushik} 
T. Kaushik {\it et al}, {\it Phys. Lett.} {\bf A 232}, 384 (1997).

\bibitem{Shioe} 
Y. Shioe {\it et al.}, {\it Il Nuovo Cimento} {\bf 112}, 1059 (1999).

\bibitem{Cardone} 
F. Cardone, A. Carpinteri, and G. Lacidogna, {\it Phys. Lett.}, {\bf A 373} (2009) 862; 
A. Carpinteri and G. Lachidogna, {\it Strain} {\bf 45}, 332 (2009).

\bibitem{Fuji} 
M. Fuji {\it et al.}, {\it Jpn. J, Appl. Phys,}  {\bf 41}, 2115 (2002).

\bibitem{LL} 
L. D. Landau and E. M. Lifshitz, {\it Electrodynamics of Continuous Media}, 
Sec. 17, Pergamon Press, Oxford (1984).

\bibitem{Fusion1} JET Team. {\it Nucl. Fusion}, {\bf 32},187 (1992)

\bibitem{Fusion2} B. E. Keen and M. L. Watkins, {\it Proceedings of the Institution of Mechanical Engineers, 
Part A: Journal of Power and Energy}, {\bf 207}, 269 (1993) 

\bibitem{Steiner} D. Steiner, {\it Proc. Int. Conf. on Nuclear Cross-sections and Technology 1975},
Eds. R. A. Schrack and C. D. Bowman, NBS Special Publication. 

\bibitem{Byrne} J. Byrne, ``Neutrons, Nuclei \& Matter'', Chapter 5.6, Dover Publications, Inc. 
Mineola, New York (2011).

\end{thebibliography}
\end{document}